\documentclass[twocolumn]{aastex631}

\usepackage{threeparttable}
\usepackage{booktabs}

\usepackage{placeins}

\usepackage{alltt}
\usepackage{ragged2e}  

\usepackage{xcolor}
\newcommand{\rev}[1]{\textcolor{black}{#1}}
\newcommand{\revb}[1]{\textcolor{black}{#1}}

\usepackage{hyperref}
\begin{document}

\title{Unveiling the Sagittarius Dwarf Spheroidal Galaxy Core with Gaia DR3: \rev{A Red Clump Distance Precise to 2\%}}

\author[0009-0004-2667-9995]{Ellie K.H. Toguchi-Tani}
\affiliation{\rev{Department of Astronomy, The University of Texas at Austin, 2515 Speedway Boulevard, Austin, TX 78712, USA}}
\affiliation{Department of Astronomy, Whitman College, 280 Boyer Ave, Walla Walla, WA 99362, USA}
\affiliation{Institute for Astronomy, University of Hawai‘i, Honolulu, HI 96822, USA}

\author[0000-0003-3244-5357]{Daniel R. Hey}
\affiliation{Institute for Astronomy, University of Hawai‘i, Honolulu, HI 96822, USA}

\author[0000-0001-5486-2747]{Thomas de Boer}
\affiliation{Institute for Astronomy, University of Hawai‘i, Honolulu, HI 96822, USA}

\author[0000-0002-0740-8346]{Peter M. Frinchaboy}
\affiliation{Department of Physics and Astronomy, Texas Christian University, 2995 S. University Dr., Fort Worth, TX 76109, USA}
\affiliation{Maunakea Spectroscopic Explorer, Canada-France-Hawaii-Telescope, Kamuela, HI 96743, USA}

\author[0000-0001-8832-4488]{Daniel Huber}
\affiliation{Institute for Astronomy, University of Hawai‘i, Honolulu, HI 96822, USA}



\begin{abstract}

The Sagittarius dwarf spheroidal galaxy provides us with the unique opportunity to study an ongoing Galactic cannibalistic event between our Milky Way Galaxy and a satellite dwarf galaxy.\rev{ Understanding this event crucially requires memberships and high-precision metallicities. Here, }we present the first major membership star catalog of the Sagittarius dwarf core ($\approx$\revb{140,000} sources) and Messier 54 ($\approx$2000 sources) with positions, proper motions, and parallaxes from $Gaia$ DR3, supplemented with metallicities from the Apache Point Observatory Galactic Evolution Experiment. We \rev{initially} isolate the Sagittarius dwarf core and Messier 54 \rev{spatially from prior} literature positions. Using evolutionary sub-samples separated within a color-magnitude diagram, we analyze the substructures of the Sagittarius core and infer its positional relationship with Messier 54 within 5D phase space. A sample of Milky Way stars from a similar galactic latitude \rev{were} used to identify contaminants and separate member stars from the core of the Sgr dSph and Messier 54 using \rev{a Gaussian Mixture Model.} We present the derived proper motion, parallaxes, and metallicities for these evolutionary sub-samples while demonstrating the precision of our sample using red clump standard candles. \rev{We find a distance modulus for the Sagittarius core and Messier 54 of $(m-M)_{0}=16.958^{+0.044}_{-0.044}$ mag and $(m-M)_{0}=16.94^{+0.047}_{-0.056}$ mag, corresponding to a heliocentric distance of $d=24.635^{+0.49}_{-0.49}$ kpc and $d=24.452^{+0.537}_{-0.602}$ kpc respectively.} With red clump distance analysis, our results imply there is no separation between the Sagittarius core and Messier 54. \rev{Finally, we describe the metallicity distributions of the evolved stars within these two systems, finding evidence for the infall scenario. }

\end{abstract}

\keywords{Sagittarius dwarf spheroidal galaxy (1423) --- Globular star clusters (656) --- Galaxy interactions (600) --- Tidal disruption(1696)}


\section{Introduction} \label{sec:Introduction}


The Sagittarius dwarf spheroidal galaxy (Sgr dSph, hereafter Sgr, \citealp{IbataSag, Ibata}), discovered nearly 30 years ago, is the closest known example of a dwarf satellite galaxy undergoing tidal disruption and accretion into a larger galaxy--the Milky Way (MW) \citep{Johnston1995, Zhao1998, Johnston1998, Helmi1999, Bellazzini2003, Putman2004, Lewis2005, BellazziniNucleus, Laporte2018, Bellazzini, Horta}. Dwarf galaxies are considered the building blocks of the hierarchical merging process \rev{\citep{GallagherDSPH, Tosi2003, Tolstoy, HelmiStreams}} and, thus, are fundamental mechanisms for forming large galaxies through a process of galactic cannibalism for spiral galaxies \citep{Hausman1978, Cannibalism}. Tidal interactions during close encounters with large host galaxies often transform star-forming, gas-rich dwarf irregular (dIrr) galaxies with small rotating disks into classical, gas-depleted, pressure-supported dwarf spheroidals \citep{Frinchaboy}. In the case of Sgr, these forces from the MW’s tidal field have elongated its core toward the Galactic plane \citep{Majewski2MASS}, complicating observations due to its proximity of $\approx6.5$ kpc below \citep{Mucciarelli2017} and $\approx15$ kpc behind the MW bulge \citep{Ibata1999}. Three pericentric passages with the MW have stripped the Sgr system, creating two massive tidal streams—a trailing arm in the south-galactic hemisphere and a leading arm in the north-galactic hemisphere--populating the MW halo with 5\% of its halo M giant stars and associated dark matter particles \citep{Majewski2MASS, Bellazzini}.

These pericentric passages and crossings through the MW disk have coincided with increased periods of star formation within Sgr and the MW, occurring 5.9, 1.9, and 1 Gyr ago \citep{deBoer, Ruiz}. These episodes have led to a dominance of intermediate-age stars within the Sgr core, aged around 4 to 6 Gyr, with metallicities ranging from [Fe/H] = $-0.4$ to $-0.6$ \citep{Siegel}. A younger\rev{, metal-rich population was first detected by \citet{Bonifacio2000} and the complex star formation history of Sgr dSph was further studied by \citet{Bonifacio2004, Sbordone2007}. This near-solar metal-rich population formed approximately 1 Gyr ago, aligns with the most recent pericentric passage, and \rev{the} metal-poor old population of [Fe/H] = $-1.41$ corresponds to pre-pericentric passages \citep{Bonifacio2004, Alfaro}}. Although material from Sgr is dispersed across a wide range of distances (10-100 kpc) due to the tidal disruption, parts of the main body remain embedded within these tidal streams, carrying stars and dark matter lost during this process \citep{Hasselquist, Fern}. While understanding the tidal streams of Sgr are crucial for probing the MW gravitational potentials and changes within the disk, understanding the core dynamics, physical properties, and kinematics, of the Sgr main body is essential for deciphering the evolution and nature of the Sgr system. Investigating these properties through galactic archaeology—studying galactic structure and evolution by examining the ages, chemical compositions, and distances of stellar populations—offers insight into Sgr and its relationship with M54. 

The main body (including the core) of Sgr also exhibits distinctive features. According to \citet{delPino}, Sgr has a bar approximately 2.5 kpc in length, inclined at $43^{\circ}\pm6^{\circ}$ in the plane of the sky, with tidal tails extending in an "S" shape from the ends of the bar. Despite significant tidal stripping, Sgr retains a minor clockwise rotation, which extends along the outer regions of its tidal stream \citep{delPino}. Currently, four globular clusters are believed to be within the main body of Sgr: Messier 54 (M54), Arp 2, Terzan 7 (Ter 7), and Terzan 8 (Ter 8) \citep{Bellazzini}. M54 (NGC 6715) lies at the photometric center of the Sgr core and has been proposed as the nucleated core of Sgr due to also lying within the densest region of the Sgr core \citep{MonacoCusp}. However, further studies \citep{BellazziniNucleus, Kunder, delPino, Zhaozhou} suggest that M54 formed independently of Sgr's nucleus, potentially being captured during tidal disruption, with \citet{Siegel2011} inferring a $\approx2$ kpc separation between the Sgr core and M54. Supporting this, \citet{Minelli} found that M54’s mean metallicity ([Fe/H] $\approx -1.30 \pm 0.12$), differing by approximately one full dex from the median metallicity of the Sgr core found in \citet{Hayes} of [Fe/H] = $-0.57$, reinforcing the idea of M54 forming independently of Sgr. Meanwhile, the central region of Sgr reveals a complex formation history, displaying a mix of young ($\leq 2.2$ Gyr), intermediate-age (around 4–6 Gyr), and old ($\geq 12.2$ Gyr) stellar populations, with metallicities spanning from $-1.41 \lessapprox$ [Fe/H] $\lessapprox +0.56$ \citep{Siegel, Alfaro, Zhaozhou}.

Following the release of $Gaia$ DR3, \citet{2021Gaia} probed the structure of the LMC and SMC, showcasing improvements from \citet{2018Gaia} where they previously determined the kinematics, including mean proper motion ($\mu_{\alpha}*$, $\mu_{\delta}$) and parallax of 75 Galactic globular clusters, the Large and Small Magellanic clouds (LMC and SMC respectively), and nine dwarf spheroidal galaxies including the Sgr core within $Gaia$ DR2. They found that photometry formerly affected by background issues caused by high stellar density in central areas of dwarfs were reduced, there was a two-fold reduction in proper motion uncertainty, and an overall increase in the precision of astrometry and photometry with significantly reduced systematic effects from $Gaia$ DR2. APOGEE DR17 \citep{APOGEEDR17} contains measurements for stars from dwarf satellite galaxies, open clusters, and globular clusters (GCs) identified from procedures detailed in \citet{Zasowski}, including Sgr and M54. The mixture of these two sky surveys allows us to constrain Sgr and M54 member stars from the MW disk to analyze the dynamics within 5D phase space. 

In this work, we describe membership selection of the Sgr core and M54 using $Gaia$ DR3 and APOGEE DR17, inspired by the methodology within \cite{2018Gaia,2021Gaia}. We describe the data selection procedure in Section~\ref{sec:Data}. In Section~\ref{sec:method} we describe the spatial separation process, removing background MW contaminants, and isolating evolutionary areas of the Sgr core and M54. In Section~\ref{sec: rc}, we use our sample of red clump (RC) stars to estimate a distance to Sgr and M54 and present our result from our distance determination. In Section~\ref{sec:Discussion}, we present results from 5D phase space analysis of the Sgr core and M54, compare our distance determinations from the RC with prior studies. We compare our membership sample to previous work and discuss the efficacy of our sample for future studies. In Section~\ref{sec:Conclusion}, we summarize our conclusions. 

\section{Data} \label{sec:Data} 

\begin{figure*}[htbp] 
    \centering
    \includegraphics[width=1\textwidth]{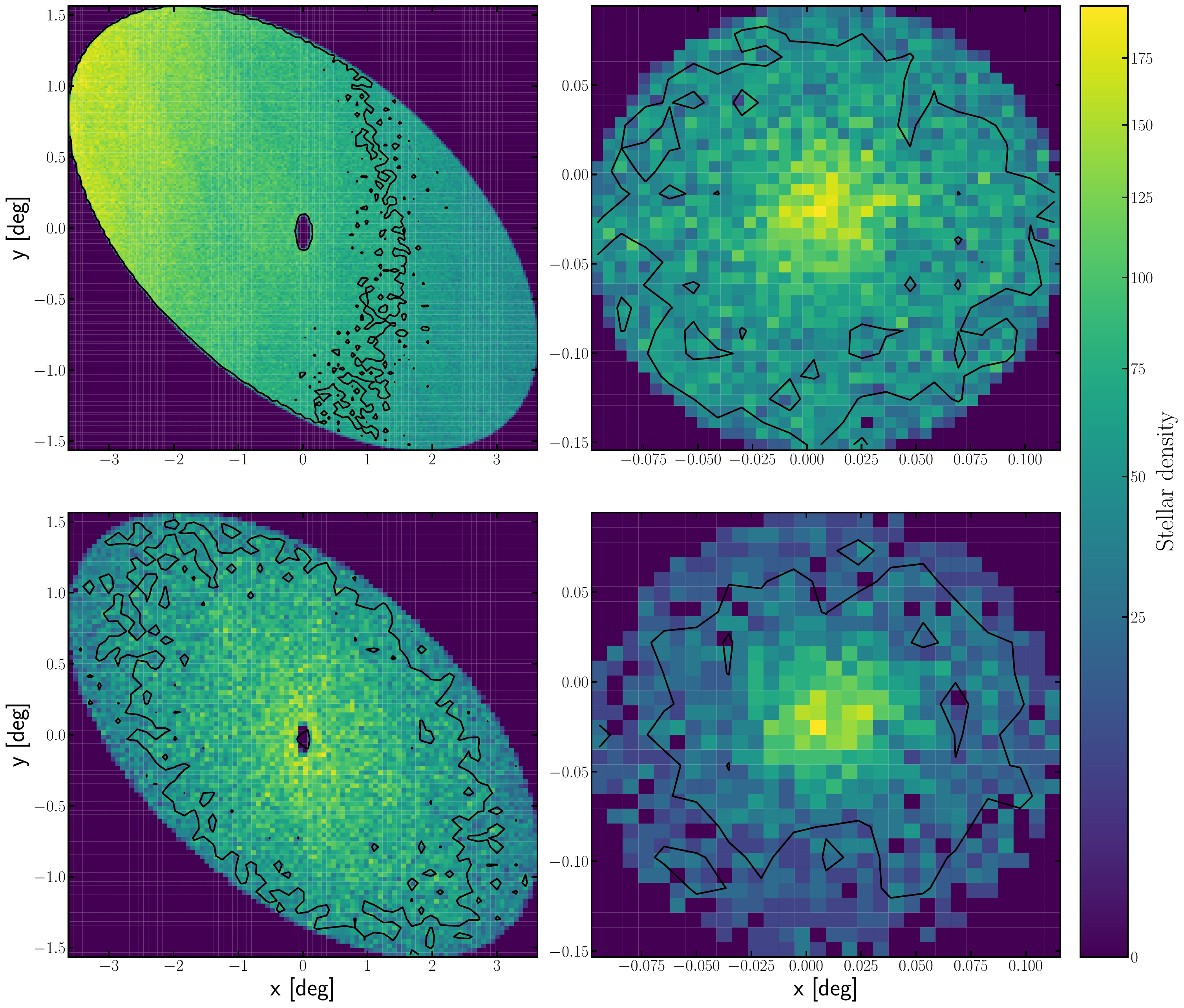}
    \caption{Sky density distribution of stars spatially selected as members of the (left) Sgr dSph core and (right) Messier 54 from $Gaia$ DR3. The black density contours present visualize the location of 80\% of stars within our sample. The bottom panels show Sgr dSph core and Messier 54 after removing Milky Way contaminants from Section~\ref{sec:mw_contam}.}
    \label{fig:density}
\end{figure*}

\subsection{Gaia DR3}

We use data from the third $Gaia$ data release ($Gaia$ DR3; \citet{Gaia2016DR3, GaiaDR3}), for the five-parameter astrometric solution; \cite{Gaia5Parameter, GaiaDR3}; right ascension ($\alpha$), declination ($\delta$), parallax ($\omega$), proper motion in right ascension direction ($\mu_{\alpha}*$; defined as $\mu_{\alpha*}=\mu_{\alpha}cos\delta$), and proper motion in declination direction ($\mu_{\delta}$). 

We extract a sample from the $Gaia$ DR3 catalog based on specific spatial criteria, where $\alpha_{\textrm{core}}, \delta_{\textrm{core}}$, are the accepted center of the Sgr core from literature in right ascension and declination respectively using the coordinates ($\alpha_{\textrm{core}}, \delta_{\textrm{core}}$) = (283.75$^{\circ}$, -30.46$^{\circ}$) \rev{\citep{Majewski2MASS}}, with a radius of $4^{\circ}$. This initial large selection radius is greater than $R_{\textrm{core}}=3.73^{\circ}$ defined in \rev{\citet{Majewski2MASS}} as the core of Sgr has a defined ellipticity of $\epsilon = 0.65 \pm 0.01$ \rev{\citep{Majewski2MASS}}. We constrain our selection to sources that have at least a five-parameter astrometric solution. This selection can be reproduced in full using the $Gaia$ ADQL query within Appendix~\ref{ADQL}. The resulting sample contains 4,378,343 objects. 

\subsection{SDSS/APOGEE DR17}

We also use data from the Apache Point Observatory Galactic Evolution Experiment (APOGEE 1 \& 2; \citealp{Majewski}), under the Sloan Digital Sky Survey III \& IV (\citealp{EisensteinSDSS, BlantonSDSS}). Observations were conducted using the 2.5 m Sloan Foundation Telescope \citep{Gunn2006} at Apache Point Observatory and the 2.5 m du Pont telescope \citep{Bowen1973} at Las Campanas Observatory, with data collected by the APOGEE-North and APOGEE-South spectrographs \citep{Wilson2019}. The selection process for APOGEE targets is described in \cite{Zasowski,Zasowski2017,Beaton2021,Santana2021}.

The APOGEE spectra were processed using the APOGEE reduction pipeline \citep{Nidever2015}, and key parameters were derived through the APOGEE Stellar Parameters and Chemical Abundances Pipeline (ASPCAP; \citealp{Garcia2016}). Further details on the spectral analysis process can be found in \citet{Shetrone2015, Smith2021}. The final data release for APOGEE-2 (DR17; \citealp{Abdurro'uf2022}) includes approximately 734,000 stars, encompassing all observations from both spectrographs collected between August 2011 and January 2021. A comprehensive overview of the quality and parameter limitations of APOGEE DR17 is available in \citet{Abdurro'uf2022}.

\section{Methodology} \label{sec:method}

\subsection{Spatial Selection} \label{sec:spatial}

From our $Gaia$ DR3 query, we compute the effective radius of the Sgr core from flattened coordinates defined in Section~\ref{FlatCoords}, using $\alpha_{\textrm{core}},\: \delta_{\textrm{core}}$, the accepted center of the Sgr core from literature (in right-ascension and declination), $\epsilon$ the ellipticity, and $PA$ the position angle, as defined from \citet{Majewski}. This spatial selection includes the globular cluster Messier 54 (M54; NGC6715), which has a tidal radius of 7.5 arcmin (0.125$^{\circ}$) \citep{M54Tidal}. 

We define our initial sample of Sgr core stars of those within the effective radius of $R_{\textrm{core}}=3.73^{\circ}$ after accounting for the ellipticity of the Sgr core, but outside of the $r_{\textrm{tidal}} = 0.125^{\circ}$ of M54. We defined our initial M54 sample with stars located within $r_{\textrm{tidal}} = 0.125^{\circ}$ of our defined $\alpha_{\textrm{core}},\: \delta_{\textrm{core}}$, along with selecting any stars outside of M54's literature radius if identified within the APOGEE membership flag (MEMBER) \citep{Zasowski, Zasowski2017}.

To refine our samples and reduce contamination from foreground Milky Way stars, we applied a modified version of the methods described by \citet{2018Gaia} and \citet{2021Gaia}. 

\begin{figure*}[htbp] 
   \centering
  \includegraphics[width=\textwidth]{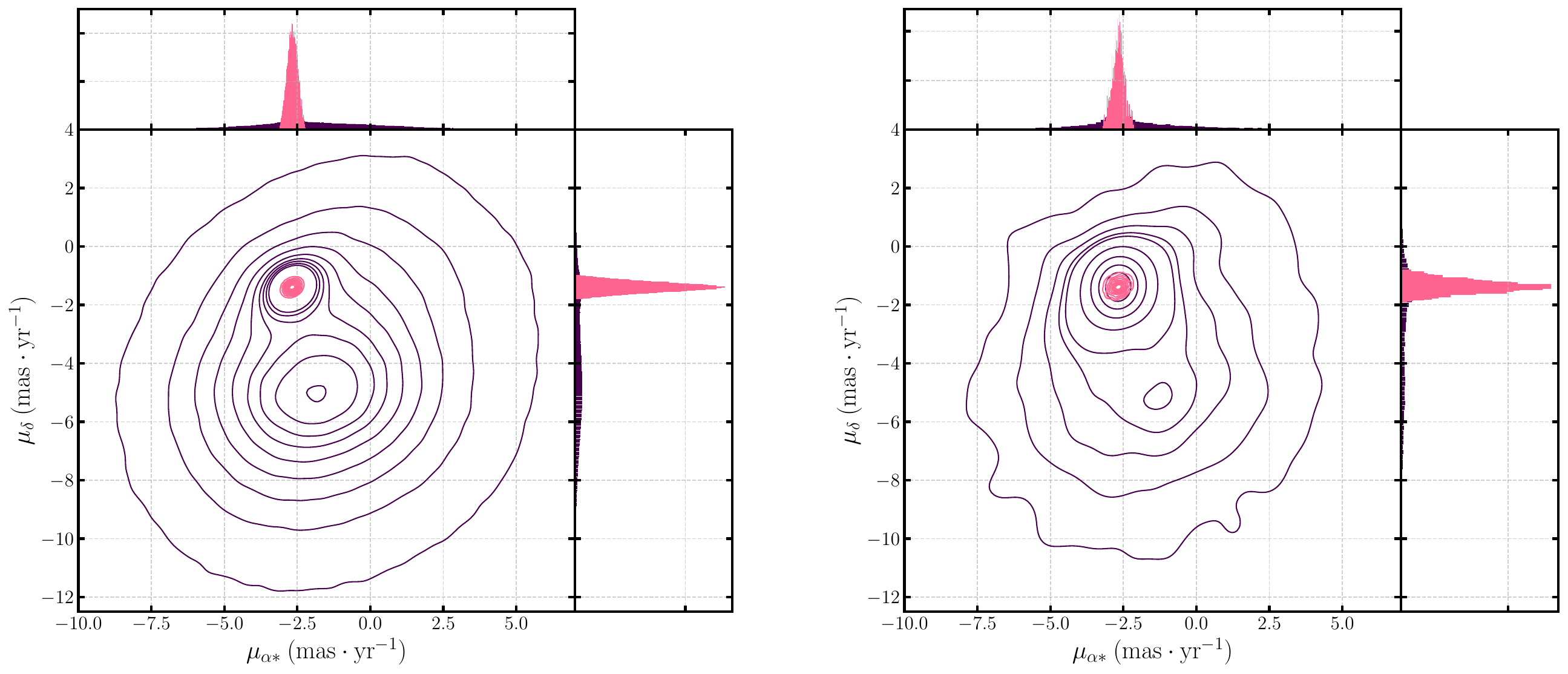}
   \caption{A bivariate contour plot of proper motion space ($\mu_\alpha*$ vs $\mu_{\delta}$) for the Sgr core (left) and M54 (right). \rev{Pink contours and histograms represent our high probability (95\%) identified members through a GMM. The purple contours and histograms represent the MW field stars.} We identify the Sgr/M54 population using literature values \rev{from \citet{2018Gaia,Zhaozhou} respectively.}}
   \label{fig:histogram}
\end{figure*}

\begin{enumerate}
    \item To ensure a high-quality five-parameter solution, we excluded stars that did not reach a minimum value of 5 within \texttt{visibility\_periods\_used}  \citep{2018Lindegren}. 
    \item We performed a cut in relative parallax error $0 < \sigma_{\omega} / \bar{\omega} < 0.5$ (which is equivalent to $\bar{\omega} - 2\sigma_{\bar{\omega}} > 0 $), which corresponds to removing stars within 5 kpc from the Sun. This eliminates stars with high-precision parallax measurements, which are more likely to be foreground objects. $Gaia$ DR3 only provides reliable distances out to $\approx 3$ kpc, literature from \citealp{MonacoDistance} puts Sgr dSph out to $\approx26$ kpc, thus we should keep large parallax errors. 
    \item We applied a limiting magnitude cut of G $<$ 20.5. This limit is introduced to remove less precise astrometry. 
    \item We applied an overall $3\sigma$ clip to the proper motion in right ascension direction ($\mu_{\alpha*}$) and declination direction ($\mu_{\delta}$) from the mean PM to further remove any outliers within our sample. 
\end{enumerate}

The resulting sample of 1,983,632 objects for Sgr core and 7,852 for M54, while cleaned of foreground sources and potential outliers, still has contaminants from Milky Way background stars. We present the sky density of our initial query sample from after the above quality cuts in the top panel for Sgr dSph and M54 stars in Figure~\ref{fig:density}. 

This contamination \rev{can be seen in \citet[Figure~3]{2018Gaia} and \citet[Figure~1]{Babusiaux2018} at around $G_{BP} - G_{RP}\approx0.5-0.7$, increasing as we move into the redder color zone. We can see this within our sample within Figure~\ref{fig:bins}.}

We reproduce the figure using our sample defined here, where background objects are also apparent in the top row of CMDs for the Sgr core (left) and M54 (right), within Figure~\ref{fig:before_after}. 


\subsection{Evolutionary Features} \label{sec: evolution}

\begin{figure*}[htbp] 
    \centering
    \includegraphics[width=0.96\textwidth]{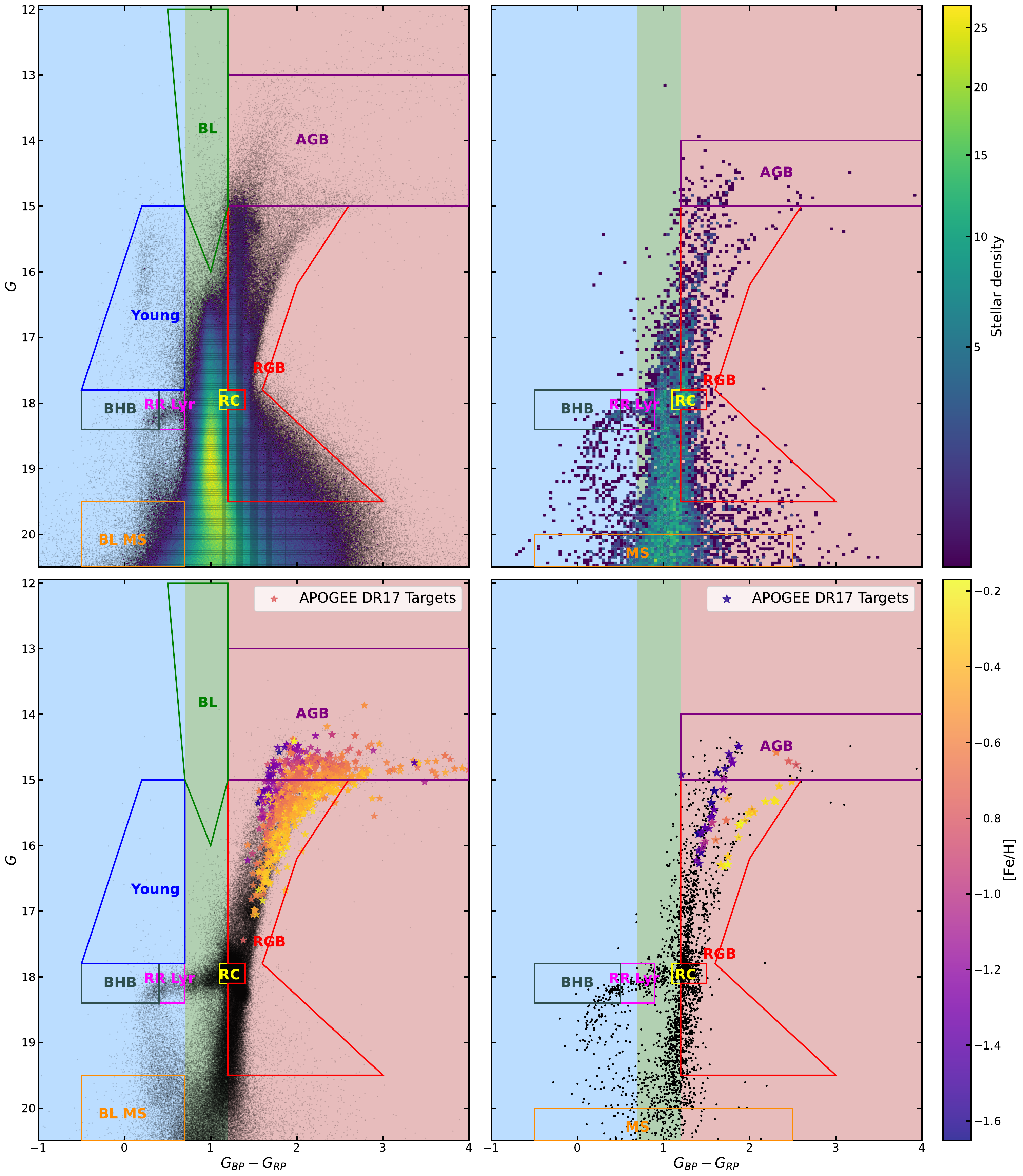}
    \caption{\rev{Color-magnitude diagrams (CMDs) for the Sgr core (left panels) and M54 (right panels), shown before (top) and after (bottom) minimizing MW contamination through a GMM \revb{ for identified evolutionary areas and an spatial over density cut} in proper motion. The background colors in all four plots indicate the blue, green, and red zones within $G_{BP}-G_{RP}$ which informed our evolutionary subsample selection detailed in \ref{sec: evolution}. The top panel points are colored by stellar density, while the multi-colored, larger, star shaped points show the APOGEE DR17 [Fe/H] values of identified member stars in the bottom panel.} Over-plotted in all panels are identified evolutionary subsamples (defined as polygons within Table~\ref{table:sgrPoly} and Table~\ref{table:m54Poly}). \rev{The black points in the bottom two panels are the high-probability identified member stars without APOGEE DR17 [Fe/H] value using a GMM.} No correction of reddening has been applied.}
    \label{fig:before_after}
\end{figure*}

Before cleaning of Milky Way contaminants, we isolate evolutionary features in our CMDs for Sgr dSph core and \rev{M54}. Based on Figure 3 in \citet{2018Gaia}, we loosely divided our CMDs into three color regions defined as $G_{BP}-G_{RP}<0.7$ dex, $0.7$ dex $< G_{BP} - G_{RP} < 1.2$ dex, and $G_{BP}-G_{RP}>1.2 $ mag (blue, green, and red regions\revb{; see Figure~\ref{fig:bins}}). The blue region appears nearly free of Milky Way contamination, whereas the green and red zones appear heavily contaminated, yet the features of the Sgr and M54 stand out, particularly when viewing stars that have measured [Fe/H] values from APOGEE DR17 in the top panel of Figure~\ref{fig:before_after}. We identify the following distinct evolutionary phases within the Sgr core based on Figure 2 from \citet{2021Gaia}: a young population extending beyond the main sequence turn-off (Young); the blue main sequence (BL MS); the red giant branch (RGB); the asymptotic giant branch (AGB), including long-period variables; the RR Lyrae (RR Lyr) region; the blue horizontal branch (BHB), which also includes RR Lyrae stars; the blue loop (BL), encompassing classical Cepheids; and the red clump (RC). Furthermore, since Sgr is an evolved, disrupted dwarf spheroidal, unlike the Large and Small Magellanic clouds in \citet{2021Gaia}, we additionally reference the \rev{evolutionary subsamples} in \rev{\citet[Figure~3]{deBoerFornax} as both Fornax and Sgr are dwarf spheroidals.}

Our defined areas are shown in Figure~\ref{fig:before_after}, with coordinates for the polygonal areas in Sgr core CMD shown in Table~\ref{table:sgrPoly} and for M54 within Table~\ref{table:m54Poly}. There are unassigned regions with the CMD diagrams, as these areas are prevalent of mixing of evolutionary areas, heavy Milky Way contamination, and/or are affected by blended stars. The most notable area of blended stars is within the green and red zones where we predict the main sequence, main sequence turn off, and sub-giant branch would lie, but cannot be disentangled. We note that the CMD for M54 excludes the region for young stars and the blue loop, as these features were not visible. The RGB and AGB within the Sgr core and M54 are noticeable features when viewing the CMDs colored by available [Fe/H] values in Figure~\ref{fig:before_after}, where both objects feature distinct RGB and AGB branches of differing metallicities (\rev{see \citet{Boecker2020} for M54 and \citet{Minelli} for Sgr dSph core}).


\begin{table}[ht]
\centering
\caption{Mean astrometry of \rev{high probability members of the} Sgr Core and Messier 54 (after position, parallax, and proper motion selection) samples divided by evolutionary phase sub-samples after cleaning of MW contaminants.}
\label{table:astrometry}

\scriptsize 
\renewcommand{\arraystretch}{1.1} 
\setlength{\tabcolsep}{3pt} 
\begin{tabular*}{\columnwidth}{@{\extracolsep{\fill}}lcccc}
\toprule
           & $\overline{\varpi}$ & $\overline{\mu_{\alpha^*}}$ & $\overline{\mu_\delta}$ & $d$ \\
\midrule
\rev{Sgr Core}   & $0.054^{+0.145}_{-0.144}$& $-2.677^{+0.181}_{-0.183}$& $-1.394^{+0.163}_{-0.162}$& $10.049^{+21.802}_{-5.886}$\\
\rev{Young} & $0.049^{+0.133}_{-0.079}$& $-2.725^{+0.176}_{-0.147}$& $-1.412^{+0.174}_{-0.127}$& $13.339^{+34.843}_{-8.049}$\\
\rev{BL MS}      & $0.052^{+0.575}_{-0.549}$& $-2.679^{+0.239}_{-0.245}$& $-1.388^{+0.217}_{-0.218}$& $2.753^{+6.963}_{-1.553}$\\
\rev{RR Lyr}  & $0.054^{+0.15}_{-0.163}$& $-2.673^{+0.2}_{-0.201}$& $-1.386^{+0.184}_{-0.175}$& $8.711^{+20.062}_{-4.541}$\\
\rev{RGB}  & $0.055^{+0.14}_{-0.134}$& $-2.678^{+0.177}_{-0.181}$& $-1.393^{+0.159}_{-0.159}$& $10.191^{+23.066}_{-6.06}$\\
\rev{AGB}  & $0.055^{+0.03}_{-0.031}$& $-2.706^{+0.143}_{-0.132}$& $-1.397^{+0.162}_{-0.159}$& $17.611^{+19.025}_{-6.13}$\\
\rev{BHB}  & $0.047^{+ 0.148}_{-0.167}$& $-2.674^{+0.195}_{-0.183}$& $-1.412^{+0.174}_{-0.127}$& $9.223^{+20.225}_{-4.866}$\\
\rev{BL}   & $0.084^{+0.089}_{-0.601}$& $-2.733^{+0.075}_{-0.042}$& $-1.411^{+0.211}_{-0.033}$& $4.785^{+25.163}_{-3.81}$\\
\rev{RC}   & $0.051^{+ 0.142}_{-0.143}$& $-2.675^{+0.182}_{-0.179}$& $-1.397 ^{+0.162}_{-0.159}$& $9.52^{+21.69}_{-4.876}$\\
\midrule
\rev{M54}        & $0.06^{+0.222}_{-0.211}$& $-2.678^{+0.205}_{-0.22}$& $-1.374^{+0.195}_{-0.178}$& $7.19^{+16.629}_{-4.541}$\\
\rev{MS}         & $0.039^{+0.658}_{-0.555}$& $-2.683^{+0.325}_{-0.338}$& $-1.35^{+ 0.247}_{-0.228}$& $2.521^{+7.023}_{-1.429}$\\
\rev{BHB} & $-0.02^{+0.207}_{-0.17}$& $-2.656^{+0.207}_{-0.239}$& $-1.368^{+ 0.158}_{-0.242}$& $8.383^{+31.615}_{-5.026}$\\
\rev{RGB}  & $0.059^{+0.129}_{-0.157}$& $-2.68^{+0.168}_{-0.183}$& $-1.384^{+ 0.154}_{-0.138}$& $10.054^{+17.927}_{-6.011}$\\
\rev{AGB} & $0.053^{+0.056}_{-0.066}$& $-2.685^{+0.104}_{-0.144}$& $-1.362^{+ 0.101}_{-0.136}$& $14.026^{+24.957}_{-6.641}$\\
\rev{RR Lyr} & $0.071^{+0.192}_{-0.204}$& $-2.717^{+0.17}_{-0.198}$& $-1.347^{+ 0.2}_{-0.219}$& $6.337^{+19.092}_{-3.362}$\\
\rev{RC}         & $0.057 ^{+0.161}_{-0.152}$& $-2.677^{+0.192}_{-0.194}$& $-1.384^{+ 0.171}_{-0.158}$& $8.612^{+21.606}_{-4.692}$\\
\bottomrule
\end{tabular*}

\vspace{1mm}
\footnotesize\noindent\textit{Note.} Parallax is in $mas$; $\mu_{\alpha^*}$ and $\mu_\delta$ in $(\mathrm{mas\cdot yr^{-1}})$; and $d$ in kiloparsecs. The zero-point was corrected for using \citealp{ZeroPoint}, yet negative parallaxes still occur. Negative parallaxes can be physically interpreted as the source going 'the wrong way around' on the sky, and are caused by parallaxes with large uncertainties \citep{NegativeParallaxes}. For parallax values present here, we adopt a 5-$\sigma$ cut to rid our mean values of outliers not fully representative of the main sample.
\end{table}


\subsection{Minimizing Milky Way Contamination} \label{sec:mw_contam}

The sample of objects obtained in the previous section of the Sgr core and M54 are still heavily contaminated by foreground and background contaminants. To clean our samples of these foreground and background contaminants, we split our larger sample into the identified evolutionary features from our contaminated CMDs in Section~\ref{sec: evolution}, along with the non-identified evolutionary areas. 

From the isolated stars within each evolutionary polygonal area in our CMDs, and all of the stars in unidentified areas we visualize two histograms: one with proper motion in the right ascension direction ($\mu_{\alpha*}$) and one in proper motion declination direction ($\mu_{\delta}$). 

In each corresponding histogram of $\mu_{\alpha}*$ and $\mu_{\delta}$ for each defined evolutionary feature and within our identified contaminated areas, there exist two distinct peaks, indicating two distinct populations within our sample as seen in Figure~\ref{fig:histogram}. Using reported literature values from \citet{2018Gaia} for the Sgr core within Table C.2, we identify one of these peaks as being Sgr member stars with a $\mu_{\alpha*} = -2.692\:\mathrm{mas\cdot yr^{-1}}$ and $\mu_{\delta} = -1.359\:\mathrm{mas\cdot yr^{-1}}$. 

To confirm that the other peak present are Milky Way contaminants, we pulled \rev{two samples from $Gaia$ DR3 at the same galactic latitude of the Sgr core, within a $4^{\circ}$ radius, but with an offset of $\pm4^{\circ}$ of Galactic longitude, where $l = 5.6^{\circ}\pm4^{\circ}, b = \rev{-}14.06^{\circ}$ (referencing $l_{\textrm{core}} = 5.6^{\circ}$ and $b_{\textrm{core}} = 14.06^{\circ}$ for the Sgr core; \rev{\citet{Majewski2MASS}}.)} We employ the same quality cuts as in Section \ref{sec:spatial}, and calculate the respective mean proper motions in the right ascension and declination direction finding, \rev{$\bar\mu_{\alpha*}= -1.811 \:\mathrm{mas\cdot yr^{-1}}$ and $\bar\mu_{\delta}= -4.808 \:\mathrm{mas\cdot yr^{-1}}$.} These calculated values align with the tallest peak within our histograms as featured in Figure~\ref{fig:histogram}. 

To remove these Milky Way contaminants, we employ \rev{a Gaussian Mixture Model (GMM) implemented using the $\texttt{Scikit-learn}$ package \citep{scikitlearn} to disentangle our member stars from each system from the MW contaminants. A GMM is a unsupervised machine learning clustering technique where the probabilistic model assumes that the data is generated from a mixture of two or more Gaussian distributions, each with its own weight, mean and variance \cite{GMM}. By providing an initial guess of the number of components, an Expectation-Maximization (EM) Algorithm can be used to find various parameters of the Gaussian components and is repeated till convergence. This also provides a probability value for each data point, dictating which Gaussian component it belongs to.}

\rev{Utilizing the literature values of the proper motion from literature values of the proper motion in the right ascension ($\mu_{\alpha*}$) and declination ($\mu_{\delta}$) direction from \citet{2018Gaia}, we apply a GMM using these values as our initial guesses for the peak of our member stars on the areas of identifiable evolutionary features within the Sgr CMD, along with defining 2 number of components. We accept any stars that report a 95\% or higher probability of belonging to the Gaussian identified belonging to Sgr. We report the FWHM values for the high probability members of the Sgr core as $FWHM_{\mu_\alpha*}=0.726$ and $FWHM_{\mu_\delta}=0.6977$. \revb{We perform a different process to areas outside identified evolutionary features, since the GMM cannot disentangle Sgr's $\mu_{\alpha*}$ and $\mu_{\delta}$ from the MW's $\mu_{\alpha*}$ and $\mu_{\delta}$ to a high probability since the MW is much more prevalent in these areas. Instead, we translate the FWHM measurements from our GMM model to an ellipse (rotated to account for Sgr's intrinsic dispersion) and spatially cut out stars that match Sgr's $\mu_{\alpha*}$ and $\mu_{\delta}$ distribution. If we were to adopt a GMM, our sample would be heavily skewed to selecting MW contamination stars, not Sgr stars. Figure~\ref{fig:histogram} visualizes the separation of Sgr-identified members (in pink) in the left-most panel and identified MW contamination (in purple). Our final sample includes 144596 identified Sgr member stars, 95199 of which are high-probability Sgr members through the GMM, which are shown in Table~\ref{tab:members}.}}

For M54, there are fewer MW background contaminants visible (\rev{due to the high density of M54$+$Sgr stars), yet we apply the same cleaning method as before for both the evolutionary subsamples and unidentified evolutionary subsamples. We use literature values from \citet{Zhaozhou} ($\bar\mu_{\alpha*}= -2.682 \:\mathrm{mas\cdot yr^{-1}}$ and $\bar\mu_{\delta}= -1.380 \:\mathrm{mas\cdot yr^{-1}}$) corroborated with the calculated $\mu_{\alpha*}$ and $\mu_\delta$ of M54 members from APOGEE DR17. These values provide an starting point for our GMM, and we again define an initial guess of 2 Gaussian components. We report the FWHM values for the high probability members of M54 as FWHM$_{\mu_\alpha*}=0.5606$ and FWHM$_{\mu_\delta}=0.5618$. We visualize the high probability identified members of M54 in the right subplot in Figure~\ref{fig:histogram}. Our final sample includes 2638 identified member stars of M54, which are shown in Table~\ref{tab:members}.}


\begin{table*}[t]
\centering
\setlength{\tabcolsep}{3pt}
\caption{Membership tables for the Sgr dSph Core and M54.}
\label{tab:members}

\begin{tabular}{lrrrccc}
\hline\hline
\multicolumn{7}{c}{\textit{Sgr dSph Sample}} \\
\hline
$Gaia$ DR3 ID & RA & Dec & $\mu_{\alpha*}$ & $\mu_{\delta}$ & $\bar{\omega}$ & GMM Probability \\
       & deg & deg & $\mathrm{mas\ yr^{-1}}$ & $\mathrm{mas\ yr^{-1}}$ & mas & \\
\hline
6736241413019624704 & 282.828419 & -31.592158 & -2.547946  & -1.413759 & 0.122520 & 0.988925 \\
6736241447378768128 & 282.840249 & -31.589329 & -2.829347  & -1.473527 & -0.084372 & 0.988629 \\
6736253060970356864 & 282.741142 & -31.566766 & -2.740621 & -1.473031 & 0.226700 & 0.990030 \\
\multicolumn{7}{c}{$\vdots$} \\
\hline\hline
\multicolumn{7}{c}{\textit{M54 Sample}} \\
\hline
$Gaia$ DR3 ID & RA & Dec & $\mu_{\alpha*}$ & $\mu_{\delta}$ & $\bar{\omega}$ & GMM Probability \\
\hline
6760412458169160320 & 283.811137 & -30.599297 & -2.431466	  & -1.168130	 & -0.123008 & 0.976058 \\
6760412462477482112 & 283.812538 & -30.600012 & -2.883681 & -1.446605 & -1.756710 & 0.979323 \\
6760412492527580032  & 283.824930 & -30.587363 & -2.463567 & -1.460903 & -0.179012 & 0.975957 \\
\multicolumn{7}{c}{$\vdots$} \\
\hline\hline
\end{tabular}

\vspace{2pt}
\footnotesize \textit{Note.} Random sample of member stars from the Sgr core and M54. \revb{Stars with a GMM Probability indicated with NaN (not a number) were selected spatially using the ellipse method.} The full version of these tables are available electronically in full online. 
\end{table*}


With our final samples for the \rev{high probability members of the} Sgr core and M54, we present an overview of the mean astrometric parameters and number of stars within each evolutionary subsample in Table~\ref{table:astrometry}. The mean parallax of the Sgr core from \citet{MonacoDistance, Minelli} (calculated from the heliocentric distance) was $\bar{\omega} \approx 0.04\:\pm0.08$ mas. Most of our parallax measurements lie somewhat near this value of $0.04\:\pm0.08$ mas, however reliable distances cannot be obtained by inverting the parallax beyond $\approx3$kpc, causing the discrepancy in our calculated distance measurements. 

We illustrate the robustness of our method of removing Milky Way contaminants by showcasing the sky density contours in the bottom panel of Figure~\ref{fig:density}. We also compare our prior CMDs before cleaning for the Sgr core and M54, before and after applying our histogram cuts in Figure~\ref{fig:before_after}.

\section{Red Clump Distance Determination} \label{sec: rc}

\begin{figure*}
    \centering
    \includegraphics[width=0.95\textwidth]{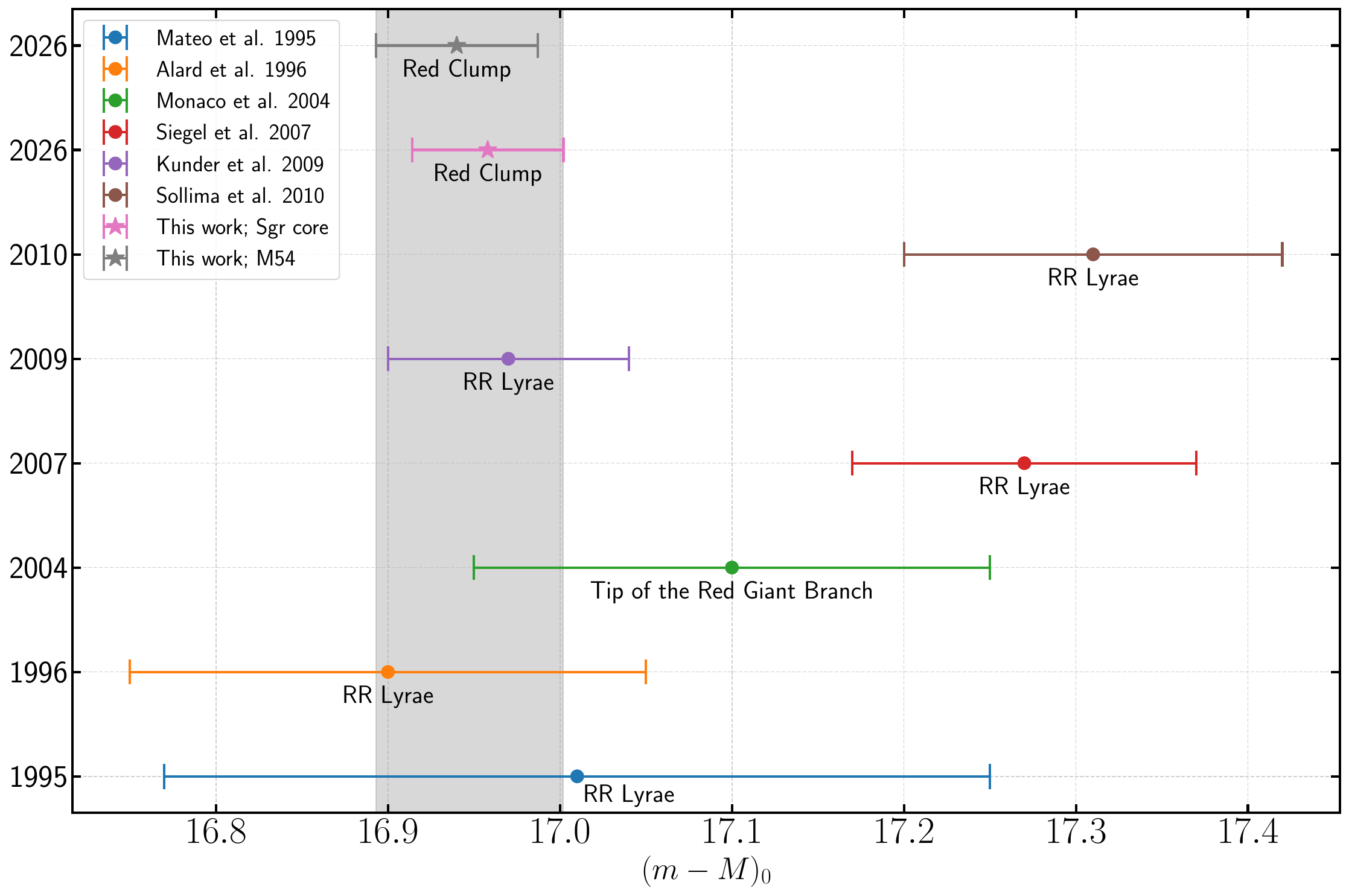}
    \caption{\rev{Distance moduli of the Sgr Core+M54 from previous works of varying distance determination methods \citep{Mateo1995,Alard,MonacoDistance,Siegel, Kunder, Sollima2010} compared to the Sgr Core and M54 distance moduli from this work.}}
    \label{fig:distanceplot}
\end{figure*}

Red clump stars are the more numerous metal-rich equivalent of horizontal branch stars, found to be standard candles in \citet{StanekandCzech}, where absolute luminosity weakly depends on age and chemical composition \citep{UdalskiRC, HelshoechtRC, GroenewegenRC, GirardiRC}. To support the precision of our membership samples for the Sgr core and M54, we perform a preliminary distance determination to these objects using RC stars as a standard candle.

We cross-match our Sgr core and M54 sample with the Two Micron All Sky Survey Point Source Catalog (2MASS; \citealp{Curti2MASS,Skrutskie2MASS}) with a matching 5 arcseconds radius to obtain $J$, $H$, and $K$-band magnitudes of the sources using the TOPCAT software \citep{TOPCAT}. 2MASS was an all-sky survey in the near-infrared; thus, cross-matching from $Gaia$, which is in the optical, to 2MASS within the NIR allows us to correct for less extinction and reddening as the $J$, $K$, and $H$-bands are less affected by interstellar dust \citep{Skrutskie2MASS}. 

We account for extinction using the $\texttt{dustmaps}$ package \citep{DustMaps}, utilizing the latest version of the three-dimensional dust map Bayestar (Bayestar19; \citealp{Green2015,Green2018,Green2019}). Bayestar19 traces dust reddening as a function of angular position and distance along a line of sight to each star. It is probabilistic, reporting how reddening increases or decreases along that sight line \citep{Green2019}. We take the mean value of the dust reddening to our Sgr core and M54 selection stars and convert the given dust reddening to extinction in 2MASS filters using the coefficients reported in Table 1 of \citet{Green2019}. We subtract the calculated extinction values from our $J$ and $K_{s}$ magnitudes to correct for extinction.

Following the methodology from \citet{StanekRC}, we create a histogram for both the Sgr core and M54, from our extinction-corrected apparent $K_{s}$ magnitudes. The peak magnitude within the distribution represents the apparent magnitude of the RC population ($m_{K_{s}}$). We use the absolute magnitude of the RC in the 2MASS band $K_{s}$ from \citet{HawkinsRC} of $-1.61 \pm 0.01$ mag to find the distance modulus $\mu$ of each system:

\[\mu_{0}=m_{K_{s}} - M_{K_{s}} - A_{K_{s}}\]

We \rev{employ} a Monte Carlo method to determine the uncertainties on our distance moduli and their corresponding distances. For each distribution of $m_{K_{s}}$ for the Sgr core and M54, we generated a KDE with a bandwidth value of 0.04. From our KDEs, we generated 10,000 resampled distributions over the range our RC was located, ending with a distribution of apparent magnitudes ($m_{K_{s}}$) for our objects. We then modeled the extinction values $A_{v}$, generating a distribution based on our results from Bayestar19, resampling 10,000 times. From the absolute magnitude of the RC in the 2MASS band from \citet{HawkinsRC}, we generate a normal distribution with a mean of $M_{K_{s}} = -1.61$ mag and a standard deviation of $0.01$ mag. We calculate the distance moduli for the Sgr core and M54 using these distributions, along with their heliocentric distances, extracting the 16th, 50th, and 84th percentiles to provide the distance moduli to the Sgr core and M54 with errors, along with their distances.

\rev{We report a final distance modulus of $\mu_{\textrm{SGR}}=16.958^{+0.044}_{-0.044}$ mag and $\mu_{\textrm{M54}}=16.94^{+0.047}_{-0.056}$ mag. These distance moduli correspond to a heliocentric distance of $d_{\textrm{SGR}}=24.635^{+0.49}_{-0.49}$ kpc for the Sgr Core and $d_{\textrm{M54}}=24.452^{+0.537}_{-0.602}$ kpc for M54.} 


\section{The Connection of the Sgr Core and M54} \label{sec:Discussion}

We isolated the Sgr core and Messier 54 spatially using $Gaia$ DR3 astrometry, applied general quality cuts from \citet{2021Gaia}, and cleaned Milky Way contaminants from our sample by identifying member stars within the histograms of proper motion in right ascension direction $\mu_{\alpha*}$ and declination direction $\mu_{\delta}$. We also calibrated distances to both of these systems using a red clump distance determination. Both distance moduli found using red clump distance methods of \rev{$\mu_{\textrm{SGR}}=16.958^{+0.044}_{-0.044}$ mag and $\mu_{\textrm{M54}}=16.94^{+0.047}_{-0.056}$ mag} fall well within literature values from \citet{Mateo1995, Alard,MonacoDistance, Siegel, Kunder,Sollima2010}, as shown in Figure~\ref{fig:distanceplot}, and prove both the precision of our sample and effectiveness of red clump distance methods. Our measured distance moduli to both systems and their respective errors overlap, suggesting that the Sgr core and M54 could occupy the same space. However, as this method only assigns one distance to the entire system–more analysis is needed. A future look into individual star distance determination methods would provide conclusive evidence if M54 is projected onto the area of Sgr's nucleus.

\begin{figure*}
    \centering
    \includegraphics[width=\textwidth]{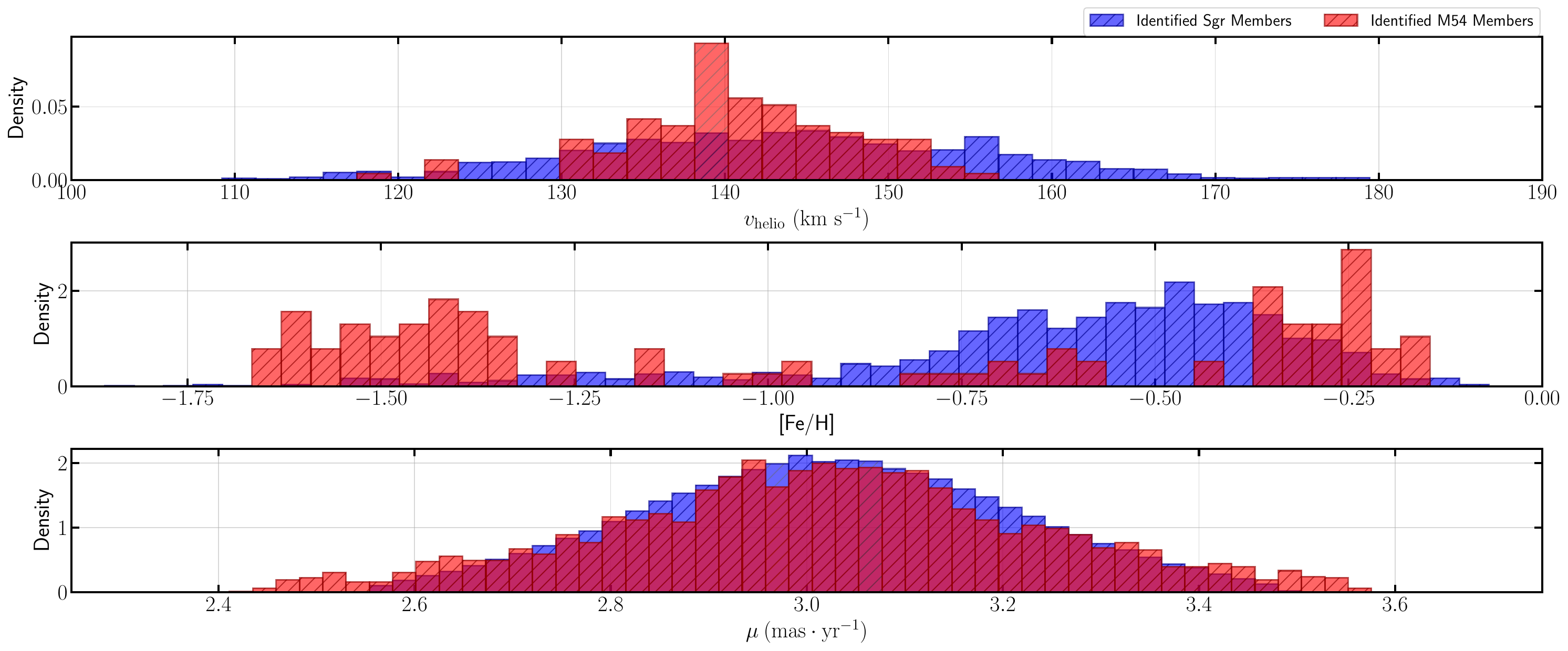}
    \caption{Measured heliocentric velocities (top panel)\rev{, metallicities ([Fe/H]; middle panel) from APOGEE DR17 for available member stars identified within this paper. Top panel is measured total proper motion ($\mu$)} from $Gaia$ DR3. Identified Sgr members are in blue; identified M54 members are in red.}
    \label{fig:radial_and_metals}
\end{figure*}

We corroborate our membership selection using available APOGEE DR17 metallicities and radial velocities from our member stars from the Sgr core and M54. From our Sgr core member selection, we have \rev{1589} stars that have radial velocities and metallicities, and from M54 we have a total of \rev{104} member stars. We show the distribution of radial velocities and metallicities in Figure~\ref{fig:radial_and_metals}. 

\rev{We report a mean radial velocity of $143.52^{+12.59}_{-12.05}\:\textrm{km}\:\textrm{s}^{-1}$ for Sgr dSph, and a mean radial velocity of $140.09^{+8.41}_{-5.06}\:\textrm{km}\:\textrm{s}^{-1}$ for M54.}  Visually, the radial velocities for the Sgr core and M54 are similar, with only $\approx2\:\textrm{km}\:\textrm{s}^{-1}$ difference in their mean. \rev{We run a Kolmogorov-Smirnov (KS) test, implemented with $\texttt{Scikit-learn}$, which determines whether two distributions are significantly one of a kind from each other and are not from the same parent distribution, on our distributions of heliocentric velocity \citep{KSTEST}. We find a moderate difference between the two distributions, and with a corresponding p-value of 0.004, this indicates that Sgr and M54's distributions of heliocentric velocity are highly unlikely to come from the same parent sample.}

\rev{We report a range of [Fe/H] of $-1.8459$ to $-0.07978$ dex for the Sgr core.} The Sgr core has a wide-range of [Fe/H] \rev{(seen visually in the bottom left CMD of Figure~\ref{fig:before_after}} with a high of nearly solar [Fe/H], to a low of $\approx-1.76$ dex, matching the range found in \citet{Siegel, Alfaro}. \rev{In the middle panel of Figure~\ref{fig:radial_and_metals}, APOGEE DR17 Sgr members are identified in blue, which form a main peak around $\textrm{[Fe/H]}\approx-0.50$ dex, then an extended tail which peaks at a lower stellar density at $\textrm{[Fe/H]}\approx-1.25$ dex. While not on a one-to-one scale, this distribution mimics the metallicity distribution found within \citet{Minelli}. \citet{Hayes} finds a dominant peak of $\textrm{[Fe/H]}\approx-0.57\:\textrm{dex}$ using APOGEE DR16, with a trailing end into the metal-poor regime down to a minimum of $\textrm{[Fe/H]}\approx-1.8\textrm\:{dex}$, which matches our findings using APOGEE DR17.}

\rev{M54 has a [Fe/H] range of $-1.633$ to $-0.169$ dex. We find two distinct peaks within the metallicity distribution, suggesting the existence of two distinct populations with M54, unlike the gradient seen in the Sgr core (see Figure~\ref{fig:radial_and_metals}.) \citet{Boecker2020}, using the Multi-Unit Spectroscopic Explorer (MUSE) spectrum, finds two dominant populations; one is old (8-14 Gyr) with a $\textrm{[Fe/H]}\approx-1.5\:\textrm{dex}$ and a younger (1 Gyr) with a $\textrm{[Fe/H]}\approx+0.25\:\textrm{dex}$. These measurements from \citet{Boecker2020} appear to line up with our samples' dominant populations within our M54 members.}

\rev{Both Sgr and M54 extend to similar ranges within total proper motion $\mu$, and initially appear to share the same distribution (bottommost panel, Figure~\ref{fig:radial_and_metals}), albeit that Sgr has more spread than M54. This is confirmed by our measured FWHMs from Section~\ref{sec:mw_contam}, where Sgr has a $FWHM\approx0.7$, while M54 has a $FWHM\approx0.56$. We do not see any major-minor axis asymmetry with these measurements. The width of these peaks suggest that the internal kinematics within each system are similar but still distinctly different. We run another KS-test to ensure that our distributions of proper motion are not from the same parent sample. We report a minor difference of 4.2\% between the two distributions, but with a corresponding p-value of 0.0002, this difference is statistically significant, meaning M54 and Sgr's proper motion distributions are highly unlikely to come from the same parent sample.}

\begin{figure*}[hbt!] 
    \centering
    \includegraphics[width=\textwidth]{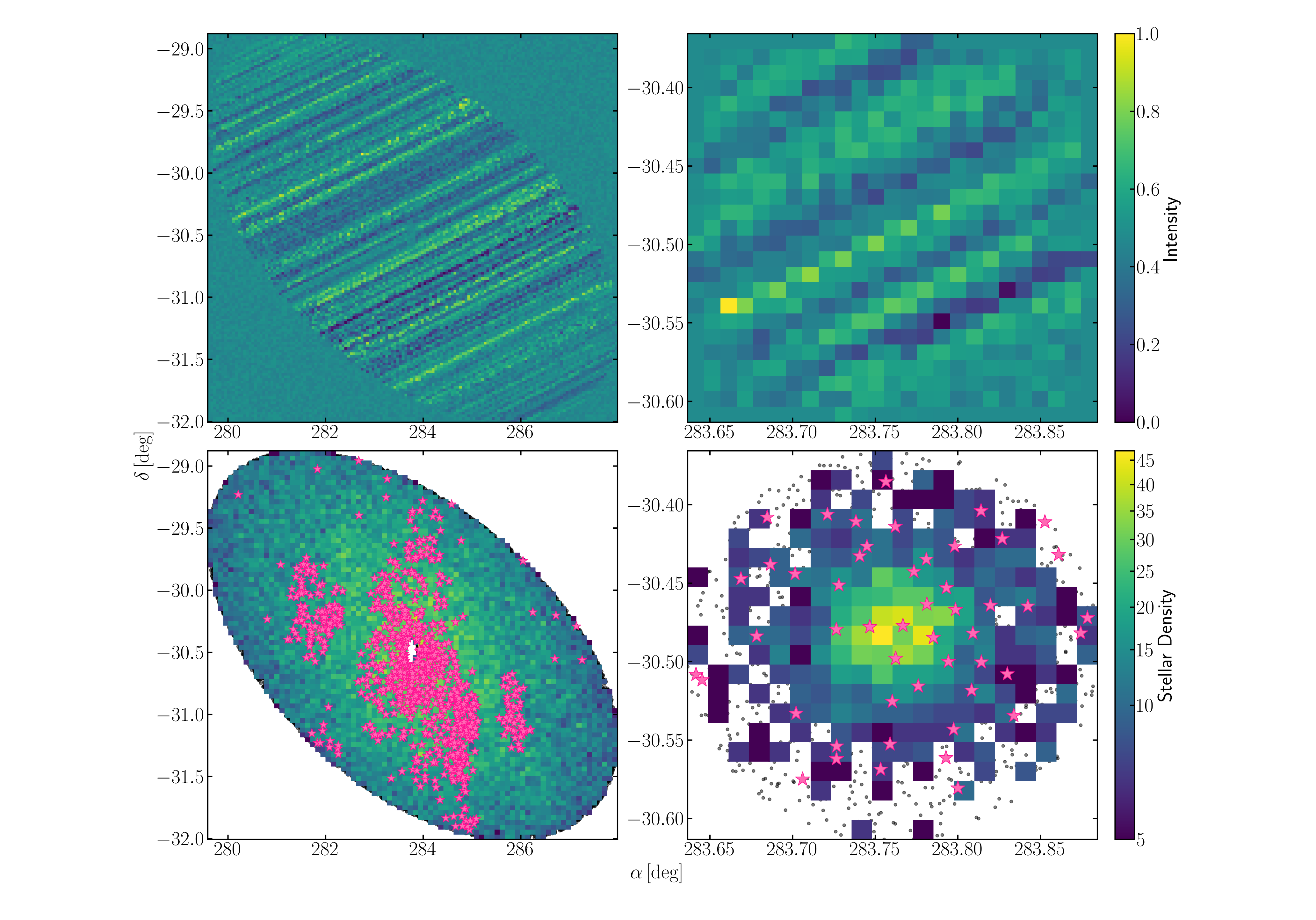}
    \caption{\rev{Illustrations of the proper motion field for the Sgr dSph core (top left) and Messier 54 (top right) using line-integral convolution (LIC). Bottom panels showcase the stellar distribution of the Sgr dSph core (bottom left) and M54 (bottom right), with APOGEE DR17 targets colored as pink stars.}}
    \label{fig:lic}
\end{figure*}

We use the line integral convolution (LIC) method, which blurs white noise textures along a vector field to visualize overall flow patterns, of the Sgr core and M54 in the right ascension direction $\mu_{\alpha*}$ and declination direction $\mu_{\delta}$. We present the visualization in Figure~\ref{fig:lic}, and notice the same dynamics in both systems–member stars concentrated towards decreasing x-direction, aligning with the pull of the leading arm of Sgr. 

The stellar density contours present in the bottom panel of Figure~\ref{fig:density} instead showcase a concentration in the central area of the Sgr core near M54, implying decoupling between the motion of the Sgr core its central concentration peak, possibly indicating that M54 remains \rev{mainly un-disrupted} despite being embedded in the Sgr core.

\rev{With the similar near-solar metallicity populations of both the Sgr and M54 system, it's possible ($\textrm{[Fe/H]}_{\textrm{Sgr}} =-0.07978\:\textrm{dex}$ and $\textrm{[Fe/H]}_{\textrm{M54}} =-0.169\:\textrm{dex}$ that this dominant near-solar stellar population within M54 is actually a subpopulation of Sgr if caused by the initial infall of M54, thus should be referred to as such. These stars should then be located near the cluster outskirts of M54. The bottom right panel of Figure~\ref{fig:lic} clearly shows an even distribution of APOGEE DR17 sources (pink stars), meaning our near-solar stellar population does not just come from the outskirt of the cluster. Sgr on the other hand, most APOGEE DR17 sources are centered nearest to where M54 has been removed from the sample, but there is still some scatter in other parts of the core.}

Evidence supporting the theory that M54 formed independently of the Sgr core comes from their respective red giant branch from their CMDs made from our sample in Figure~\ref{fig:before_after}. In the bottom panels, where the RGB and AGB branches are indicated with APOGEE DR17 [Fe/H], the Sgr core displays a rough gradient along its RGB and corresponding AGB branches, marked by three different metallicities. This lines up with the three pericentric interactions with the core of the Milky Way that kick-started star formation in both galaxies \citep{deBoer, Ruiz}. The CMD of M54 lacks these three distinct RGB and AGB branches, instead bearing two distinct branches as found in Figure 2 of \citet{Mucciarelli2017}. This suggests that M54 only went through one or two pericentric passages, enriching the cluster, kick-starting star formation, implying M54 infell during the first passage of Sgr \rev{, which is reflected in the findings of two dominant stellar populations in \citet{Boecker2020}}. Similarly, the density contours present in the bottom right panel of Figure~\ref{fig:density} and the jump from the uniform center to a scattered value outside the first density contour lines with the central density cusp found in \citet{MonacoCusp} validate our sample further. 

The most recent paper for a Sgr core membership selection was \citet{Minelli}, where they identified 450 member stars in $Gaia$ Early Data Release 3 (EDR3). Cross-matching with our sample, we recover \rev{447} stars from Minelli's sample. From those \rev{447} stars, none have [Fe/H] measurements from APOGEE DR17, and there are no $Gaia$ DR3 [M/H] measurements using General Stellar Parametrizer from Spectroscopy (GSP-Spec) for our sample. There are [M/H] measurements for these 427 within $Gaia$ DR3 General Stellar Parametrizer from Photometry (GSP-Phot), however these values exhibit substantial biases compared to literature values and are only useful at a qualitative level especially for red giant branch stars with low quality parallax measurements \citep{GSPPHOT}.  Other attempts to gain [Fe/H] measurements from The Large Sky Area Multi-Object Fiber Spectroscopic Telescope (LAMOST) and The GALactic Archaeology with HERMES (GALAH) fail to capture Sgr dSph core and M54 member stars within our sample, unfortunately not allowing us a direct comparison of \citealp{Minelli} [Fe/H] measurements taken from the Fibre Large Array Multi Element Spectrograph (FLAMES) on the Very Large Telescope (VLT).  


\section{Conclusion} \label{sec:Conclusion}
In this paper, we have used $Gaia$ DR3 and APOGEE DR17 to produce the largest catalog of stars from the Sgr Core and Messier 54, separating member stars from the heavily contaminated area from foreground and background Milky Way stars. We present the precision of our samples using red clump distance determinations using magnitudes from 2MASS. Our main conclusions are as follows:

\begin{itemize}
    \item We demonstrate the strength of $Gaia$ DR3 astrometry combined with APOGEE DR17 spectra to separate the Sgr core with M54, along with a multi-Gaussian fit to reduce MW contaminants for systems near the disk. We have a membership catalog of \revb{144596 total} stars in the Sgr core and \rev{2638} in M54. 
    \item We measure the distance to both the Sgr core and M54 using red clump distancing methods from a cross-matched sample with 2MASS, finding a heliocentric distance of \rev{$d_{\textrm{SGR}}=24.635^{+0.49}_{-0.49}$ kpc for the Sgr Core and $d_{\textrm{M54}}=24.452^{+0.537}_{-0.602}$ kpc for M54.} Our distance measurements agree with \rev{prior} literature but are more precise due to the larger member sample identified in our work \rev{and a new RC distance determination method.}
    \item We analyze the motion, density contours, and [Fe/H] of the Sgr core and M54 to understand whether M54 formed independently and was captured during a passage, or is the Sgr nucleus. Preliminary distance measurements and the dynamics of our sample imply they co-exist, but the chemical abundance \rev{history} suggest the systems formed independently.
\end{itemize}

The results presented here provide the first glimpse into the full dynamics of the Sgr core and its relationship to M54 in $Gaia$ DR3. Utilizing this large membership sample and cross-matching with current and future all-sky ground-based and space-based surveys will allow further study into the dynamics of the Sgr core and M54. Combined with masses and ages from asteroseismic data sets, we can further understand the formation, and relationship of the Sgr core and M54. 

\section{Acknowledgments} \label{sec:Acknowledgements}

E.K.H.T.T. acknowledges support from the Research Experience for Undergraduates program at the Institute for Astronomy, University of Hawai‘i-Mānoa funded through the National Science Foundation grant $\#$6109694. E.K.H.T.T. also acknowledges support from the Whitman Internship Grant through Whitman College. E.K.H.T.T. would like to thank the Institute for Astronomy for their hospitality during this project. 

E.K.H.T.T. also gratefully acknowledges support for this research from the National Science Foundation’s Research Experience for Undergraduates program PHY-2244258 at Texas Christian University. 

E.K.H.T.T. would like to additionally extend gratitude to Dr. Robyn Sanderson from the University of Pennsylvania, Dr. Sukanya Chakrabarti from the University of Alabama, Huntsville, Dr. Marc Hon from the Massachusetts Institute of Technology, Dr. Andrea Dobson from Whitman College, Dr. Matthew Shetrone from the University of Californa, Santa Cruz, Aldo \rev{Sepulveda} from the University of Texas at Austin School of Law, \rev{and Dr. Keith Hawkins from the University of Texas at Austin} \revb{and Dr. Eugene Vasiliev from the University of Surrey} for their insightful conversations and feedback on this work. 

D.H. acknowledges support from the Alfred P. Sloan Foundation and the National Science Foundation (AST-2009828). 

P.M.F. acknowledges support from the National Science Foundation (AST-2206541).

\revb{We also thank the reviewer for their helpful feedback and comments to improve this work.}

This work has made use of data from the European Space Agency (ESA) mission
{\it Gaia} (\url{https://www.cosmos.esa.int/gaia}), processed by the {\it Gaia}
Data Processing and Analysis Consortium (DPAC,
\url{https://www.cosmos.esa.int/web/gaia/dpac/consortium}). Funding for the DPAC
has been provided by national institutions, in particular the institutions
participating in the {\it Gaia} Multilateral Agreement.

Funding for the Sloan Digital Sky Survey IV has been provided by the Alfred P. Sloan Foundation, the U.S. Department of Energy Office of Science, and the Participating Institutions. 

SDSS-IV acknowledges support and resources from the Center for High Performance Computing at the University of Utah. The SDSS website is www.sdss4.org.

SDSS-IV is managed by the Astrophysical Research Consortium for the Participating Institutions of the SDSS Collaboration including the Brazilian Participation Group, the Carnegie Institution for Science, Carnegie Mellon University, Center for Astrophysics | Harvard \& Smithsonian, the Chilean Participation Group, the French Participation Group, Instituto de Astrof\'isica de Canarias, The Johns Hopkins University, Kavli Institute for the Physics and Mathematics of the Universe (IPMU) / University of Tokyo, the Korean Participation Group, Lawrence Berkeley National Laboratory, Leibniz Institut f\"ur Astrophysik Potsdam (AIP),  Max-Planck-Institut f\"ur Astronomie (MPIA Heidelberg), Max-Planck-Institut f\"ur Astrophysik (MPA Garching), Max-Planck-Institut f\"ur Extraterrestrische Physik (MPE), National Astronomical Observatories of China, New Mexico State University, New York University, University of Notre Dame, Observat\'ario Nacional / MCTI, The Ohio State University, Pennsylvania State University, Shanghai Astronomical Observatory, United Kingdom Participation Group, Universidad Nacional Aut\'onoma de M\'exico, University of Arizona, University of Colorado Boulder, University of Oxford, University of Portsmouth, University of Utah, University of Virginia, University of Washington, University of Wisconsin, Vanderbilt University, and Yale University.

%



\software{astropy \citep{2013A&A...558A..33A,2018AJ....156..123A},  
          dustmaps \citep{DustMaps}, 
          numpy \citep{harris2020array},
          scipy \citep{2020SciPy-NMeth},
          matplotlib \citep{Hunter:2007},
          lic \citep{lic},
          scikit-learn \citep{scikitlearn}
          }



\bibliography{sample631}{}
\bibliographystyle{aasjournal}


\clearpage

\appendix \label{appendix}

\section{Gaia DR3 ADQL Queries}\label{ADQL}

The following $Gaia$ DR AQDL \rev{query produces} our spatial sample for the Sgr dSph core in full.

\begin{alltt}
SELECT
  gaia_source.designation, gaia_source.source_id, gaia_source.ra, gaia_source.ra_error,
  gaia_source.dec, gaia_source.dec_error, gaia_source.parallax, gaia_source.parallax_error,
  gaia_source.parallax_over_error, gaia_source.pm, gaia_source.pmra, gaia_source.pmra_error,
  gaia_source.pmdec, gaia_source.pmdec_error, gaia_source.visibility_periods_used,
  gaia_source.phot_g_mean_mag, gaia_source.bp_rp, gaia_source.radial_velocity,
  gaia_source.radial_velocity_error, gaia_source.rv_template_fe_h,
  gaia_source.phot_variable_flag, gaia_source.l, gaia_source.b, gaia_source.non_single_star,
  gaia_source.has_epoch_rv, gaia_source.teff_gspphot, gaia_source.logg_gspphot,
  gaia_source.mh_gspphot, gaia_source.distance_gspphot, gaia_source.ag_gspphot,
  gaia_source.ebpminrp_gspphot
FROM gaiadr3.gaia_source
WHERE CONTAINS(
  POINT('ICRS', gaiadr3.gaia_source.ra, gaiadr3.gaia_source.dec),
  CIRCLE('ICRS',
    COORD1(EPOCH_PROP_POS(283.8292,-30.5453,0,-2.6500,-.8800,140.0000,2000,2016.0)),
    COORD2(EPOCH_PROP_POS(283.8292,-30.5453,0,-2.6500,-.8800,140.0000,2000,2016.0)), 4)
)=1
AND gaia_source.phot_g_mean_mag IS NOT NULL
AND gaia_source.parallax IS NOT NULL
AND gaia_source.bp_rp IS NOT NULL
\end{alltt}

\FloatBarrier
\section{Flattened Coordinates} \label{FlatCoords}

Below we showcase the equations to compute the flattened coordinates for the Sgr core and account of the elliptical shape of the core.\rev{The basic principle of the flattened coordinates is to set up a new reference frame centered on the object of study. We also provide definitions for each of the variables: right ascension of the Sgr core ($\alpha_{\textrm{core}}$), declination of the Sgr core ($\delta_{\textrm{core}}$), tangent-plane coordinate along the increasing right-ascension direction ($x_{i}$), tangent-plane coordinate along the increasing declination direction ($x_{n}$), position angle of the major axis of the Sgr core ($PA$), rotation angle used to align the coordinate system with the major axis of the Sgr core ($pa$), rotated tangent-plane coordinate aligned with the major axis of the Sgr core ($x_{i, new}$), rotated tangent-plane coordinate aligned with the minor axis ($x_{n, new}$), ellipticity ($e$), and elliptical radius from the center of the Sgr core in the flattened coordinate system ($r$).}

\begin{equation} \label{eq1}
x_i = \frac{sin(\alpha-\alpha_\textrm{core})}{sin(\delta_\textrm{core})\cdot\tan(\delta)+\cos(\delta_\textrm{core})\cdot\cos(\alpha-\alpha_\textrm{core})}
\end{equation}

\begin{equation} \label{eq2}
x_n = \frac{\cos(\delta_\textrm{core})\cdot\tan(\delta)-sin(\delta_\textrm{core})\cdot \cos(\alpha-\alpha_\textrm{core})}{sin(\delta_\textrm{core})\cdot \tan(\delta)+\cos(\delta_\textrm{core})\cdot\cos(\alpha-\alpha_\textrm{core})}
\end{equation}

\begin{equation} \label{eq3}
pa = 90 - PA
\end{equation}

\begin{equation} \label{eq4}
x_{i\_ \text{new}} = x_i \cdot \cos(pa) + x_n \cdot \sin(pa)
\end{equation}

\begin{equation} \label{eq5}
x_{n\_ \text{new}}= \frac{-x_i\cdot \sin(pa)+x_n \cdot \cos(pa)}{1-e}
\end{equation}

\begin{equation} \label{eq6}
r = \sqrt{x_{i\_ \text{new}}^2+x_{n\_ \text{new}}^2}
\end{equation}


\newpage
\section{\rev{Polygon Coordinates}}\label{Coordinates}
\FloatBarrier
\begin{deluxetable}{ll}
\tablecaption{Polygon coordinates for identified evolutionary features in the Sgr dSph core CMD within $Gaia$ DR3 magnitudes\label{table:sgrPoly}, adapted from CMDs within \rev{\citet{2021Gaia}.}}
\tablehead{
  \colhead{\textbf{Evolutionary Feature}} & \colhead{\textbf{Polygon Coordinates}}
}
\startdata
Young   & [0.7, 17.8], [-0.5, 17.8], [0.2, 15], [0.7, 15], [0.7, 17.8] \\
BL MS   & [-0.5, 20.5], [-0.5, 19.5], [0.7, 19.5], [0.7, 20.5], [-0.5, 20.5] \\
RGB     & [1.2, 19.5], [1.2, 18.1], [1.4, 18.1], [1.4, 17.8], [1.2, 17.8], [1.2, 15], \\
        & [2.6, 15], [2.0, 16.2], [1.6, 17.8], [3.0, 19.5] \\
AGB     & [1.2, 15], [1.2, 13], [1.5, 13], [4.0, 13], [4.0, 15], [1.2, 15] \\
RR Lyr  & [0.4, 17.8], [0.4, 18.4], [0.7, 18.4], [0.7, 17.8], [0.4, 17.8] \\
BHB     & [-0.5, 17.8], [-0.5, 18.4], [0.4, 18.4], [0.4, 18.4], [0.4, 17.8], [-0.5, 17.8] \\
BL      & [1, 16], [0.7, 15], [0.5, 12], [1.2, 12], [1.2, 13], [1.2, 15] \\
RC      & [1.1, 18.1], [1.1, 17.8], [1.4, 17.8], [1.4, 18.1], [1.1, 18.1] \\
\enddata
\end{deluxetable}

\begin{deluxetable}{ll}
\tablecaption{Polygon coordinates for identified evolutionary features in the M54 CMD within $Gaia$ DR3 magnitudes\label{table:m54Poly}, adapted from CMDs within \rev{\citet{2021Gaia}.}}
\tablehead{
  \colhead{\textbf{Evolutionary Feature}} & \colhead{\textbf{Polygon Coordinates}}
}
\startdata
MS      & [-0.5, 20.5], [-0.5, 20], [2.5, 20], [2.5, 20.5], [-0.5, 20.5] \\
RGB     & [1.2, 19.5], [1.2, 18.1], [1.5, 18.1], [1.5, 17.8], [1.2, 17.8], [1.2, 15], \\
        & [2.6, 15], [2, 16.2], [1.6, 17.8], [3, 19.5] \\
AGB     & [1.2, 15], [1.2, 14], [1.5, 14], [4.4, 14], [4.4, 15], [1.2, 15] \\
RR Lyr  & [0.5, 17.8], [0.5, 18.4], [0.9, 18.4], [0.9, 17.8], [0.5, 17.8] \\
BHB     & [-0.5, 17.8], [-0.5, 18.4], [0.5, 18.4], [0.5, 18.4], [0.5, 17.8], [-0.5, 17.8] \\
RC      & [1.1, 18.1], [1.1, 17.8], [1.5, 17.8], [1.5, 18.1], [1.1, 18.1] \\
\enddata
\end{deluxetable}


\newpage
\FloatBarrier
\section{\rev{Gaia BP-RP Contamination}}\label{Color Bin Width}
\begin{figure}
    \centering
    \includegraphics[width=0.5\columnwidth]{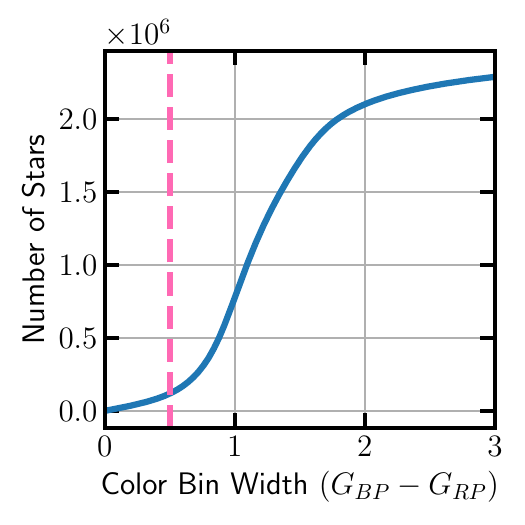}
    \caption{\rev{Number of stars as a function of color bin width in $(G_{BP}-G_{RP})$. The vertical dashed line indicates our chosen division of the beginning of uptick of MW contamination (green area in our CMDs) at $G_{BP}-G_{RP}\approx0.5 - 0.7.$}}
    \label{fig:bins}
\end{figure}

\FloatBarrier
\section{Gaia DR3 Isochrones}

\begin{figure}
    \centering
    \includegraphics[width=0.9\textwidth]{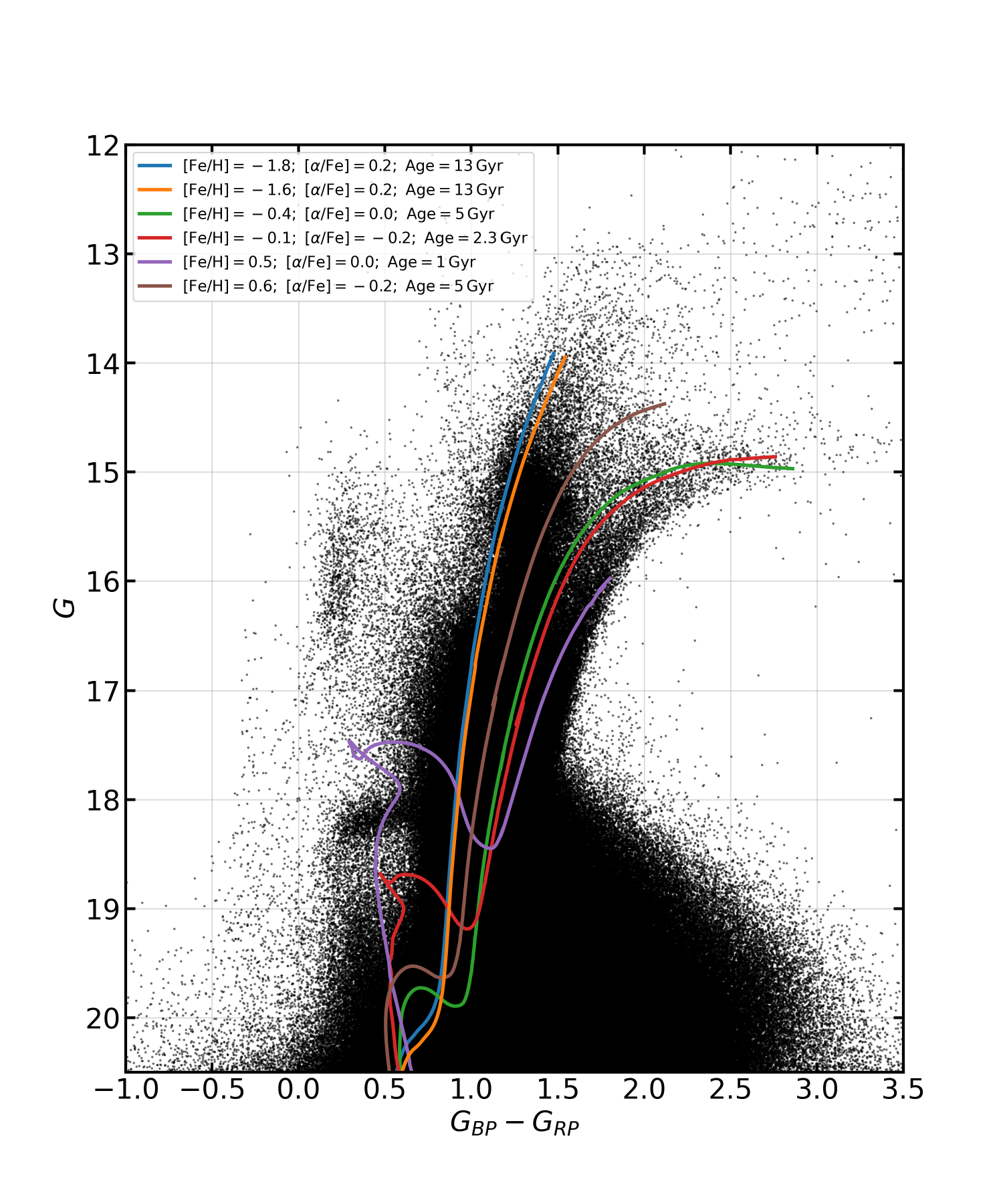}
    \caption{\revb{Isochrones from the Dartmouth Stellar Evolution Program (DSEP; \citet{Dartmouth2008}), using age, [Fe/H], and [$\alpha$/Fe] values from \citet{Siegel, Alfaro} over plotted on our full MW contaminated Sgr sample. Most of these isochrones trace out most of our identified evolutionary areas, except for the blue loop, and young area.}}
    \label{fig:isochrones}
\end{figure}



\end{document}